\documentclass[format=sigconf,nonacm]{acmart}

\usepackage{amsmath}
\usepackage{xspace}
\usepackage{multirow}
\usepackage[linesnumbered,ruled,vlined]{algorithm2e}
\SetKwInput{KwRem}{Remarks}
\usepackage{paralist}

\newcommand{\Dfn}[1]{\textbf{\emph{#1}}}
\newcommand{\Code}[1]{\texttt{\small #1}}
\newcommand{\Dom}[1]{\ensuremath{\mathit{dom}(#1)}\xspace}

\newcommand{\WFSteps}{\ensuremath{S}\xspace}
\newcommand{\WFPrec}{\ensuremath{\preceq}\xspace}
\newcommand{\WFPrecNE}{\ensuremath{\prec}\xspace}
\newcommand{\aWAP}{\ensuremath{W}\xspace}
\newcommand{\WFSchema}[5]{\ensuremath{(#1, #2, #3, #4, #5)}}
\newcommand{\WFUsers}{\ensuremath{U}\xspace}
\newcommand{\WFSAuth}{\ensuremath{A}\xspace}
\newcommand{\WFConst}{\ensuremath{C}\xspace}
\newcommand{\aUser}{\ensuremath{u}\xspace}
\newcommand{\aStep}{\ensuremath{s}\xspace}
\newcommand{\WFConstraint}[2]{\ensuremath{(#1, #2)}}
\newcommand{\aStepSet}{\ensuremath{T}\xspace}
\newcommand{\aRemSet}{\ensuremath{\Delta}\xspace}
\newcommand{\aPPlanSet}{\ensuremath{\Theta}\xspace}
\newcommand{\aPPlan}{\ensuremath{\theta}\xspace}

\newcommand{\aSeq}{\ensuremath{\tau}\xspace}
\newcommand{\anotherSeq}{\ensuremath{\pi}\xspace}

\newcommand{\SOD}[2]{\ensuremath{\mathsf{sod}(#1, #2)}}
\newcommand{\BOD}[2]{\ensuremath{\mathsf{bod}(#1, #2)}}
\newcommand{\ENT}[3]{\ensuremath{\mathsf{ent}(#1, #2, #3)}}
\newcommand{\aRel}{\ensuremath{\rho}\xspace}

\newcommand{\WFRem}[2]{\ensuremath{#1\langle #2\rangle}}

\newcommand{\NP}{\ensuremath{\mathsf{NP}}\xspace}
\newcommand{\PSPACE}{\ensuremath{\mathsf{PSPACE}}\xspace}
\newcommand{\SigmaTwo}{\ensuremath{\Sigma_2^p}\xspace}
\newcommand{\SigmaThree}{\ensuremath{\Sigma_3^p}\xspace}
\newcommand{\PiTwo}{\ensuremath{\Pi_2^p}\xspace}

\newcommand{\SAT}{\ensuremath{\mathit{SAT}}\xspace}
\newcommand{\WSP}{\ensuremath{\mathit{WSP}}\xspace}
\newcommand{\SRCP}{\ensuremath{\mathit{SRCP}}\xspace}
\newcommand{\ORCP}{\ensuremath{\mathit{ORCP}}\xspace}
\newcommand{\CRCP}{\ensuremath{\mathit{CRCP}}\xspace}
\newcommand{\DRCP}{\ensuremath{\mathit{DRCP}}\xspace}
\newcommand{\DHC}{\ensuremath{\mathit{DHC}}\xspace}
\newcommand{\SuccRad}[1]{\text{\textit{SUCCINCT}-$#1$-\textit{RADIUS}\xspace}}

\newcommand{\aspIf}{\ensuremath{\operatorname{:-}}}
\newcommand{\aspNot}{\ensuremath{\operatorname{not}}}

\begin{document}


\title{Results in Workflow Resiliency}
\subtitle{Complexity, New Formulation, and ASP Encoding}
\author{Philip W.L.~Fong}
\affiliation{University of Calgary}
\email{pwlfong@ucalgary.ca}

\begin{abstract}
  First proposed by Wang and Li in 2007, workflow resiliency is a
  policy analysis for ensuring that, even when an adversarial
  environment removes a subset of workers from service, a workflow can
  still be instantiated to satisfy all the security constraints.  Wang
  and Li proposed three notions of workflow resiliency: static,
  decremental, and dynamic resiliency.  While decremental and dynamic
  resiliency are both \PSPACE-complete, Wang and Li did not provide a
  matching lower and upper bound for the complexity of static
  resiliency.

  The present work begins with proving that static resiliency is
  \PiTwo-complete, thereby bridging a long-standing complexity gap in
  the literature.  In addition, a fourth notion of workflow
  resiliency, one-shot resiliency, is proposed and shown to remain in
  the third level of the polynomial hierarchy.  This shows that
  sophisticated notions of workflow resiliency need not be
  \PSPACE-complete.  Lastly, we demonstrate how to reduce static and
  one-shot resiliency to Answer Set Programming (ASP), a modern
  constraint-solving technology that can be used for solving reasoning
  tasks in the lower levels of the polynomial hierarchy.  In summary,
  this work demonstrates the value of focusing on notions of workflow
  resiliency that reside in the lower levels of the polynomial
  hierarchy.
\end{abstract}

\maketitle

\section{Introduction}

Confidentiality, integrity, and availability are three essential goals
of security. This paper is about availability considerations in
workflow authorization models.

A workflow is an application-level abstraction of a business process.
Access control in a workflow application is captured in a workflow
authorization model \cite{ThomasSandhu, AtluriHuang, atluri,
  Tan-etal:2004, CramptonMonitor}, the main idea of which is to
realize the Principle of Least Privilege \cite{LeastPrivilege} through
permission abstraction \cite{Baldwin:1990}.  Instead of granting
permissions directly to users, permissions are granted to the steps of
a workflow. When a user is assigned to perform a step in the workflow,
the permissions are then made available to the user.  Two additional
access control features are typically found in a workflow
authorization model. First, qualification requirements can be imposed
on each workflow step. An example is to require that the ``Prepare
Budget'' step be carried out by an ``Account Clerk,'' while the
``Account Review'' step be carried out by an ``Account Manager''
(example taken from \cite{Wang-Li:2010}).  This is typically framed in
terms of roles in an underlying Role-Based Access Control (RBAC)
model.  Second, further security constraints may be imposed across
steps to prevent abuse. A classical example would be
separation-of-duty constraints: e.g., the ``Create Payment'' step and
the ``Approve Payment'' step must be carried out by two distinct users
(example taken from \cite{Crampton-etal:2013}). Recent works in
workflow authorization models have considered arbitrary binary
constraints, thereby introducing into workflow authorization models
an element of Relationship-Based Access Control (ReBAC)
\cite{Wang-Li:2007, Wang-Li:2010, Khan-Fong:2012}.

As permissions are now encapsulated in workflow steps, we want to make
sure that the workflow can actually be executed, or else the situation
amounts to a denial of service.  In other words, one must ensure that
it is possible to assign users to workflow steps, so that all
qualification requirements and security constraints are satisfied.
This notion of availability has been known in the literature as the
\Dfn{workflow satisfiability problem (WSP)} \cite{atluri,
  CramptonMonitor, Wang-Li:2010, Crampton-etal:2013}.  WSP can be used
as a policy analysis to help the workflow developer debug her
formulation, so that the latter is not overly constrained.

A major landmark in the study of the WSP has been the work of Wang and
Li \cite{Wang-Li:2007, Wang-Li:2010}, who first introduced into the
literature the application of Fixed Parameter Tractable (FPT)
algorithms \cite{Downey-Fellows:2013} to solve WSP.  Subsequently,
major breakthroughs in the design of FPT algorithms for WSP have been
invented by the research group at Royal Holloway University of London,
including kernelization algorithms \cite{Crampton-etal:2012,
  Crampton-etal:2013, Crampton-Gutin:2013:SACMAT}, exploitation of
problem symmetry in pattern-based algorithms \cite{Cohen-etal:2014,
  Cohen-etal:2016}, and novel problem formulations
\cite{Crampton-etal:2015:SACMAT, Crampton-etal:2017:SACMAT},

A second major contribution of Wang and Li's work is the introduction
of an advanced notion of availability that is stronger than workflow
satisfiability. That notion is workflow resiliency \cite{Wang-Li:2007,
  Wang-Li:2010}, the subject of this paper.  The idea is to anticipate
catastrophes that may remove users from service. One would like to
ensure that, even if the adversarial environment has taken away a
certain number of users, the workflow is still satisfiable by the
remaining personnel.  More specifically, Wang and Li proposed three
different notions of workflow resiliency, namely, static, decremental,
and dynamic resiliency.  While decremental and dynamic resiliency have
been shown to be \PSPACE-complete, Wang and Li did not provide a
matching upper and lower bound for the complexity of static
resiliency: static resiliency is in \PiTwo and is \NP-hard. Whether
static resiliency is \PiTwo-complete has remained an open problem
since the notion was first conceived a decade ago.  Bridging this
complexity gap is the first motivation of this work.

The last few years have witnessed a renewed interest in the study of
workflow resiliency \cite{Mace-etal:2014, Mace-etal:2015:QEST,
  Mace-etal:2015:SERENE, Crampton-etal:2017:JCS}.  Two lines of
research have been representative.  The first is the work of Mace
\emph{et al.} \cite{Mace-etal:2014, Mace-etal:2015:QEST,
  Mace-etal:2015:SERENE}, who argue that it is more important to
assess to what (quantitative) degree of resiliency a workflow enjoys,
rather than to test if the workflow is resilient or not (binary).  The
goal is to offer guidance to the workflow developer in terms of
refining the formulation of the business process.

While Mace \emph{et al.}'s approach is valuable, it is the position of
this paper that deep insights of the workflow can be gained by
evaluating the workflow against multiple and incomparable notions of
workflow resiliency: e.g., statically resilient for a budget of $t_1$
and decrementally resilient for a budget of $t_2$, where $t_2 < t_1$.
In fact, each notion of workflow resiliency captures a family of
attack scenarios. Confirming that a workflow is resilient in terms of
several incomparable notions of resiliency offers deep insight into
the formulation of the workflow.  We therefore need a good number of
notions of workflow resiliency, rather than just a few.  More than
that, we need notions of workflow resiliency that are not
computationally prohibitive to test.  Unfortunately, a pessimistic
reader of Wang and Li may come to the (unwarranted) conclusion that
static resiliency is an exception rather than a rule, and that most
notions of workflow resiliency are \PSPACE-complete.  This pessimism
is understandable as the notions of workflow resiliency proposed by
Wang and Li are formulated in terms of strategic games. If resiliency
is fundamentally a \PSPACE-complete phenomenon, then hoping for an
efficient solution may be unrealistic.  A second motivation of this
work is to demonstrate that there are indeed useful notions of
workflow resiliency that are not as prohibitive as decremental and
dynamic resiliency.

A second line of recent work in workflow resiliency is that of
Crampton \emph{et al.} \cite{Crampton-etal:2017:JCS}, who devised a
first FPT algorithm for deciding dynamic resiliency.  The parameter
they used was $k + t$, where $k$ is the number of steps in the
workflow, and $t$ is the number of users that the adversary can
remove.  While $k$ is universally accepted to be a small parameter in
the literature \cite{Wang-Li:2010, Crampton-etal:2013}, $t$ is not.
For example, if $t$ is a fixed fraction of the number of users (e.g.,
5\% of the user population), it already grows much faster than is
acceptable for an FPT algorithm.  It is the position of this paper
that parameterizing the problem using $t$ is not fruitful (see \S
\ref{sec-conclusion}, however, for an example of adversary models in
which such a parameterization could make sense).  In this light,
non-FPT approaches are still valuable when $t$ is not a small
parameter. The formulation of algorithmic solutions for workflow
resiliency without assuming a small $t$ is the third motivation of
this work.

This paper has three contributions:
\begin{enumerate}
\item In \S \ref{sec-srcp}, static resiliency is proven to be
  \PiTwo-complete, thereby bridging the long-standing complexity gap
  in the work of Wang and Li \cite{Wang-Li:2007, Wang-Li:2010}.  This
  result also provides the intellectual justification for the third
  contribution below.
\item In \S \ref{sec-orcp}, we dispel the pessimistic interpretation
  of Wang and Li's work by formulating a new notion of workflow
  resiliency, one-shot resiliency, which is more sophisticated than
  static resiliency, and nevertheless remains in the third level of
  the polynomial hierarchy (\SigmaThree-complete).  This means that
  useful notions of workflow resiliency can be formulated without
  flirting with \PSPACE-completeness.
\item We advocate the use of Answer Set Programming (ASP)
  \cite{ASP-Book, Gelfond-Lifschitz:1988}, a modern constraint-solving
  technology, for deciding static and one-shot resiliency.  ASP has
  been shown to be particularly fitted for reasoning problems in the
  lower levels of the polynomial hierarchy \cite{Eiter-etal:1995,
    Eiter-etal:2007, Sakama-Rienstra:2017, Brewka-etal:2017}. In \S
  \ref{sec-asp-encoding}, we demonstrate the feasibility of this
  approach by presenting ASP encodings of static and one-shot
  resiliency. These reductions employ an encoding technique known as
  the model saturation technique \cite{Eiter-etal:1995,
    Eiter-etal:2007}.  This solution approach does not require the
  parameter $t$ to be small.
\end{enumerate}

\section{Background: Workflow Satisfiability and Resiliency}

This section provides a brief introduction to workflow satisfiability
and resiliency, in order to prepare the reader to understand the rest
of this paper.  All the materials presented in this section have
already appeared in previous work (particularly \cite{Wang-Li:2007,
  Wang-Li:2010}).

\subsection{Workflow Satisfiability}
A workflow is the abstract representation of a business process.
\begin{definition}
  A \Dfn{workflow} $(\WFSteps, \WFPrec)$ is a partial ordering of steps.
  For steps $\aStep_1, \aStep_2 \in \WFSteps$, we write $\aStep_1
  \WFPrecNE \aStep_2$ whenever $\aStep_1 \WFPrec \aStep_2$ but
  $\aStep_1 \neq \aStep_2$.
\end{definition}
Steps are tasks to be executed by users, and the partial ordering
expresses the causal dependencies among steps. If two steps are
ordered by \WFPrecNE, then they must be executed in that order;
otherwise, they can interleave in any arbitrary manner.

When a workflow is executed, users are assigned to the steps,
sometimes in an incremental manner.
\begin{definition}
  Given a workflow $(\WFSteps, \WFPrec)$ and a set \WFUsers of users,
  a \Dfn{partial plan} is a function
  $\aPPlan : \aStepSet \rightarrow \WFUsers$ such that (a)
  $\aStepSet \subseteq \WFSteps$, and (b) \aPPlan is \Dfn{causally
    closed}: that is, for $\aStep_1, \aStep_2 \in \WFSteps$, if
  $\aStep_1 \in \aStepSet$ and $\aStep_2 \WFPrecNE \aStep_1$, then
  $\aStep_2 \in \aStepSet$.  A partial plan \aPPlan is also called a
  \Dfn{plan} when $\Dom{ \aPPlan } = \WFSteps$.
\end{definition}
Security constraints, such as seperation of duty, may be imposed on a
workflow in order to prevent abuse.  
\begin{definition}[in the style of \cite{Cohen-etal:2014, Cohen-etal:2016}]
  A \Dfn{workflow authorization policy} \aWAP is a 5-tuple \WFSchema{
    \WFSteps }{ \WFPrec }{ \WFUsers }{ \WFSAuth }{ \WFConst }, where
  the components are defined as follows:
  \begin{itemize}
  \item $(\WFSteps, \WFPrec)$ is a workflow.
  \item \WFUsers is a set of users.
  \item $\WFSAuth \subseteq \WFSteps \times \WFUsers$ is the \Dfn{step
      authorization policy}, which lists for each step those users who
    are qualified to carry out the step.
  \item $\WFConst$ is a set of \Dfn{constraints}. Each constraint has
    the form \WFConstraint{ \aStepSet }{ \aPPlanSet }. The set
    $\aStepSet \subseteq \WFSteps$ specifies the steps constrained by
    the constraint.  The component \aPPlanSet is a set of partial
    plans, each with \aStepSet as its domain.  The set \aPPlanSet
    specifies the combinations of assignments that are permitted by
    the constraint.
  \end{itemize}
\end{definition}

The following definition specifies when a (partial) plan satisfies the
requirements imposed by a workflow authorization policy.
\begin{definition}
  Suppose \aPPlan is a partial plan for the workflow authorization
  policy
  $\aWAP = \WFSchema{ \WFSteps }{ \WFPrec }{ \WFUsers }{ \WFSAuth }{
    \WFConst }$. We say that \aPPlan is \Dfn{valid} if and only if the
  following conditions hold:
  \begin{itemize}
  \item For every step $\aStep \in \WFSteps$,
    $(\aStep, \aPPlan ( \aStep )) \in \WFSAuth$.
    \item For every constraint $\WFConstraint{ \aStepSet }{ \aPPlanSet
      } \in \WFConst$, if $\aStepSet \subseteq \Dom{ \aPPlan }$, then
      there exists $\aPPlan' \in \aPPlanSet$ such that
      for every $\aStep \in \aStepSet$, $\aPPlan(\aStep) =
      \aPPlan(\aStep)'$. 
    \end{itemize}
    A plan is valid if and only if it is valid as a partial plan.
  \end{definition}
  The following are some examples of constraints. Following
  \cite{Wang-Li:2007, Wang-Li:2010}, we focus mostly on entailment
  constraints, including the generalization by
  \cite{Crampton-etal:2012, Crampton-etal:2013}. Our ASP encodings of
  \S \ref{sec-asp-encoding} can easily handle the cardinality
  constraints of \cite{Crampton-etal:2012, Crampton-etal:2013} as
  well.
\begin{example}
  Suppose \WFSteps is a set of steps, and \WFUsers is a set of users.
  \begin{itemize}
  \item Suppose $\WFSteps_1$ and $\WFSteps_2$ are non-empty subsets of
    \WFSteps, and $\aRel \subseteq \WFUsers \times \WFUsers$ is a
    binary relation.
    We write \ENT{ \aRel }{ \WFSteps_1 }{ \WFSteps_2 } to denote the
    \Dfn{entailment constraint} $\WFConstraint{ \aStepSet }{
      \aPPlanSet }$ for which $\aStepSet = \aStepSet_1 \cup \aStepSet_2$,
    and $\aPPlanSet = \{ \aPPlan \in \WFUsers^\aStepSet \mid \exists
    \aStep_1 \in \aStepSet_1, \aStep_2 \in \aStepSet_2 \,.\,
    (\aPPlan( \aStep_1 ), \aPPlan( \aStep_2 )) \in \aRel \}$.
  \item Crampton \emph{et al.} classify entailment constraints in to
    various types \cite{Crampton-etal:2012, Crampton-etal:2013}. A
    \Dfn{type-1} entailment constraint is one in which both step sets
    are singleton sets. We overload notation and write \ENT{ \aRel }{
      \aStep_1 }{ \aStep_2 } to denote \ENT{ \aRel }{ \{\aStep_1\} }{
      \{\aStep_2\} }.  A \Dfn{type-2} entailment constraint is one in
    which exactly one of the two step sets is a singleton set. A
    \Dfn{type-3} entailment constraint is one in which neither of the
    step sets is a singleton set.
  \item The \Dfn{separation-of-duty constraint} \SOD{ \aStep_1 }{
      \aStep_2 } is the type-1 entailment constraint \ENT{ \neq }{
      \aStep_1 }{ \aStep_2 }.  Similarly, the \Dfn{binding-of-duty
      constraint} \BOD{ \aStep_1 }{ \aStep_2 } is defined to be \ENT{
      = }{ \aStep_1 }{ \aStep_2 }.
  \end{itemize}
\end{example}
 When one formulates a workflow authorization policy, one must ensure
that the constraints are not overly restrictive to the point that no
valid plan exists.
\begin{definition}
  A workflow authorization policy \aWAP is \Dfn{satisfiable} if and
  only if at least one valid plan exists.  \WSP is the language of
  workflow authorization policies that are satisfiable.
\end{definition}

\begin{theorem}[\cite{Wang-Li:2007, Wang-Li:2010}]
  \WSP is \NP-complete.
\end{theorem}
Even though \WSP is theoretically intractable, previous work has
demonstrated that it can be decided with moderate efficiency by apply
modern \SAT solving technologies \cite{Wang-Li:2010}, and with
even greater efficiency by Fixed-Parameter Tractable algorithms
\cite{Wang-Li:2010, Crampton-etal:2013, Cohen-etal:2014, Cohen-etal:2016}.

\subsection{Workflow Resiliency}

For mission critical business processes, a degree of availability
higher than workflow satisfiability is often desired. The basic idea
is to anticipate catastrophic events, which may render some users
unavailable for duty.  The goal is to ensure that there is enough
redundancy in human resources so that the workflow can execute to
completion even when accidents occur.

The first notion of workflow resiliency models a workflow that runs
for a very short period of time (e.g., in minutes).  Some users become
unavailable prior to workflow execution.  Due to the short duration of
the workflow, no further users are removed from service.
\begin{definition}[Static Resiliency \cite{Wang-Li:2007, Wang-Li:2010}]
  A workflow authorization policy
  $\aWAP = \WFSchema{ \WFSteps }{ \WFPrec }{ \WFUsers }{ \WFSAuth }{
    \WFConst }$ is \Dfn{statically resilient} for an integer budget
  $t \geq 0$ if and only if, for every subset \aRemSet of \WFUsers
  that has size $t$ or less, there is a valid plan
  $\aPPlan : \WFSteps \rightarrow (\WFUsers \setminus \aRemSet)$ that
  does not assign the users in \aRemSet.
\end{definition}
Workflow resiliency is typically described in terms of two-person
games: Player 1 attempts to construct a valid plan, while Player 2,
who models the adversarial environment, counters Player 1 by removing
users from service.  Static resiliency can thus be modelled by a
two-person game that is played in one round: Player 2 first removes up
to $t$ users, and then Player 1 constructs a valid plan with the
remaining users. A workflow authorization policy is statically
resilient when Player 1 can win no matter how Player 2 plays.

The next notion of workflow resiliency models the situation in which
the workflow runs for a moderately long time (e.g., within a day).  During
the execution of the workflow, more and more users become unavailable.
\begin{definition}[Decremental Resiliency \cite{Wang-Li:2007, Wang-Li:2010}]
  A workflow authorization policy
  $\aWAP = \WFSchema{ \WFSteps }{ \WFPrec }{ \WFUsers }{ \WFSAuth }{
    \WFConst }$ is \Dfn{decrementally resilient} for integer budget
  $t \geq 0$ if and only if Player 1 can win the
  \Dfn{decremental resiliency game} no matter how Player 2 plays.

  The decremental resiliency game is a two-player game that proceeds
  in rounds.  At any time, the configuration of the game is a pair
  $(\aRemSet, \aPPlan)$, where $\aRemSet \subseteq \WFUsers$, and
  $\aPPlan$ is a partial plan.  In the initial configuration, both
  \aRemSet and \aPPlan are $\emptyset$.

  Each round begins by Player 2 choosing some users from \WFUsers to
  be added to \aRemSet, so long as $|\aRemSet| \leq t$. (A legitimate
  special case is when no user is chosen. Also, users added to
  \aRemSet remain there until the end of the game.) Next, Player 1
  extends \aPPlan by assigning a user from
  $\WFUsers \setminus \aRemSet$ to a not-yet-assigned step.

  Player 2 wins right away if $\WFUsers \setminus \aRemSet$ becomes
  empty, or if \aPPlan becomes invalid.  Player 1 wins if \aPPlan is
  eventually turned into a valid plan (i.e., every step is assigned a
  user).
\end{definition}

The third notion of workflow resiliency models situations in which the
workflow runs for an extended period of time (e.g., in days). Once an
accident occurs to remove some users from service, they return to work
before the next accident has a chance to occur.
\begin{definition}[Dynamic resiliency \cite{Wang-Li:2007, Wang-Li:2010}]
  A workflow authorization policy
  $\aWAP = \WFSchema{ \WFSteps }{ \WFPrec }{ \WFUsers }{ \WFSAuth }{
    \WFConst }$ is \Dfn{dynamically resilient} for integer budget
  $t \geq 0$ if and only if Player 1 can win the \Dfn{dynamic
    resiliency game} no matter how Player 2 plays.

  The dynamic resiliency game is a two-player game that proceeds in
  rounds.  At any time, the configuration of the game is a partial
  plan \aPPlan.  Initially, \aPPlan is $\emptyset$.

  Each round begins by Player 2 choosing a subset \aRemSet of \WFUsers
  such that $|\aRemSet| \leq t$.  Next, Player 1 extends \aPPlan by
  assigning a member of $\WFUsers \setminus \aRemSet$ to a
  not-yet-assigned step.

  Player 2 wins right away if $\aRemSet = \WFUsers$, or if \aPPlan
  becomes invalid.  Player 1 wins if \aPPlan is eventually turned into
  a valid plan.
\end{definition}

We write \SRCP for the language of pairs $(\aWAP, t)$ for which \aWAP
is statically resilient for a budget $t$. Similarly, we write \CRCP
and \DRCP for the respective language of decremental and dynamic
resiliency.

The three notions of resiliency are totally ordered in terms of how
demanding they are.
\begin{theorem}[\cite{Wang-Li:2007, Wang-Li:2010}] 
  $\DRCP \subset \CRCP \subset \SRCP$.
\end{theorem}
Note the proper set inclusion in the statement above.  While \SRCP is
the least demanding notion of resiliency among the three, its
computational complexity is also the least intimidating.
\begin{theorem}[\cite{Wang-Li:2007, Wang-Li:2010}] 
  \label{thm-complexity}
  \DRCP and \CRCP are \PSPACE-complete.
  \SRCP is in \PiTwo, and is \NP-hard.
\end{theorem}
Note that the upper and lower bound for \SRCP do not match.  Since the
publication of Wang and Li's works a decade ago, whether \SRCP is
\PiTwo-complete has remained an open problem.  The starting point of
the present work is to provide an affirmative answer to this problem.

\section{Static Resiliency Revisited}
\label{sec-srcp}

The first main contribution of this work is the following result:
\begin{theorem} \label{thm-srcp-pi-two-hard}
  \SRCP is \PiTwo-hard.
\end{theorem}
In addition to providing a matching lower bound for Wang and Li's
upper bound (Theorem \ref{thm-complexity}), this result has practical implications for the choice of
solution strategy for \SRCP.  \PiTwo-completeness implies that it is
unlikely one can employ \SAT-solving technologies to solve \SRCP.
One is now driven to employ constraint-solving technologies
that are designed for problems in the second level of the polynomial
hierarchy.  As we shall see in \S \ref{sec-asp}, one such technology
is Answer-Set Programming.

\begin{proof}
  We sketch a polynomial-time Karp reduction from the \PiTwo-complete
  problem, \Dfn{Dynamic Hamiltonian Circuit (\DHC)} \cite{DHC,
    PHCatalog}, to \SRCP.  
\begin{description}
\item[Problem:] \DHC \cite{DHC}
\item[Instance:] A simple graph\footnote{An undirected graph is
    \Dfn{simple} if it contains neither loops nor multi-edges.} $G =
  (V, E)$, and an edge set $B \subseteq E$.
\item[Question:] Is it the case that for every $D \subseteq B$ with
  $|D| \leq |B|/2$, $G_D$ has a Hamilton cycle?
\item[Remark:] The graph $G_D$ is defined to be $(V, E \setminus D)$.
\end{description}

The proposed reduction takes as input an instance of \DHC, which
consists of a graph $G = (V, E)$ and a set $B \subseteq E$.  The
reduction returns an instance of \SRCP consisting of a budget
$t = |B|/2$ and a workflow authorization policy \WFSchema{ \WFSteps }{
  \WFPrec }{ \WFUsers }{ \WFSAuth }{ \WFConst }:

\begin{enumerate}
\item $\WFSteps = \mathit{SV} \cup \mathit{SE}$
  is a set of $2N$ steps, where $N = |V|$:
  \begin{align*}
    \mathit{SV} &= \{ \mathit{sv}_1, \mathit{sv}_2,
                  \ldots, \mathit{sv}_N \} \\
    \mathit{SE} &= \{ \mathit{se}_1, \mathit{se}_2,
    \ldots, \mathit{se}_N \}
  \end{align*}
  These steps model a Hamiltonian cycle: each step in
  $\mathit{SV}$ models a vertex in the Hamiltonian cycle, and each
  step in
  $\mathit{SE}$ corresponds to an edge in the
  Hamiltonian cycle.
\item The partial order \WFPrec is simply the equality relation ($=$).
  In other words, steps can be executed in any order.
\item $\WFUsers = \mathit{UV} \cup \mathit{UE}$
  is
  the set of users defined as follows:
  \begin{align}
\mathit{UV} & = \bigcup_{v \in V} \mathit{UV}_v \nonumber\\
\mathit{UE} &= \bigcup_{e \in E} \mathit{UE}_e \nonumber\\
 \mathit{UV}_v & = \{ (v, 1), (v, 2), \ldots, (v, t+1) \}
  & &  \text{for $v \in V$} \label{eqn-SRCP-vertex}\\
 \mathit{UE}_e & = \{ (e, 1), (e, 2), \ldots, (e,
  t+1) \} & & \text{for $e \in E \setminus B$} \label{eqn-SRCP-edge-not-B}\\
\mathit{UE}_e & = \{ (e, 1) \} & & \text{for $e \in
  B$} \label{eqn-SRCP-edge-in-B}
  \end{align}
  Intuitively, the users represent vertices and edges in graph $G$.  A
  plan, which assigns users to steps, effectively identifies vertices
  and edges that participate in the Hamiltonian cycle.  All vertices
  and those edges \emph{not} in $B$ have $t+1$ copies (see
  \eqref{eqn-SRCP-vertex} and \eqref{eqn-SRCP-edge-not-B}). That means
  the adversary cannot make these vertices and edges unavailable by
  removing $t$ users.  There are, however, only one copy of those
  edges in $B$ (see \eqref{eqn-SRCP-edge-in-B}).  The adversary can
  prevent these edges from participating in the Hamiltonian cycle.
\item The step authorization policy \WFSAuth ensures that users
  representing vertices are assigned to steps representing vertices,
  and the same for edges.
\[
   \WFSAuth = (\mathit{SV} \times \mathit{UV}) 
      \cup (\mathit{SE} \times \mathit{UE})
\]
\item The constraint set
  $\WFConst = \WFConst_\mathit{cir} \cup \WFConst_\mathit{ham}$ is
  made up of two types of constraint.
  Intuitively, the constraints in $\WFConst_\mathit{cir}$ ensure that
  each valid plan identifies a circuit of size $N = |V|$, while the
  constraints in $\WFConst_\mathit{ham}$ ensure that the identified
  circuit is Hamiltonian.
  \begin{enumerate}
  \item $\WFConst_\mathit{cir}$ is a set of type-1 entailment
  constraints induced by the binary relation $\mathit{incident}$.  For
  every edge $e \in E$ connecting vertices $u,v \in V$, define a
  binary relation
  $\mathit{incident}_e \subseteq \WFUsers \times \WFUsers$:
\[
   \mathit{incident}_e = 
      \left( \mathit{UE}_e \times \mathit{UV}_u
      \right) \cup
      \left( \mathit{UE}_e \times \mathit{UV}_v
      \right) 
\]
 In other words, $\mathit{incident}_e$ relates a user representing $e$
to a user representing one of the two vertices connected by $e$.
Now, $\mathit{incident}$ is defined as follows:
\[
   \mathit{incident} = \bigcup_{e \in E} \mathit{incident}_e
\]
  We also write $\mathit{incident}^{-1}$ to represent the
  converse\footnote{The converse of $R \subseteq X \times Y$ is the
    relation $\{ (y, x) \in Y \times X \mid (x, y) \in R \}$.}
  of $\mathit{incident}$. In short, $\mathit{incident}^{-1}$ relates a
  user representing a vertex to a user representing an edge that has
  that vertex as one of its two ends.
  Now, define $\WFConst_{\mathit{cir}} = \WFConst_1 \cup \WFConst_2
  \cup \WFConst_3$:
  \begin{align*}
    \WFConst_1 &= \{ \ENT{ \mathit{incident}^{-1} }{
      \mathit{sv}_i }{ \mathit{se}_i } \mid 1 \leq i
      \leq N \}\\
    \WFConst_2 &= \{ \ENT{ \mathit{incident} }{
      \mathit{se}_i }{ \mathit{sv}_{i+1} } \mid 1 \leq i
      \leq N-1 \}\\
    \WFConst_3 &= \{ \ENT{ \mathit{incident} }{
      \mathit{se}_N }{ \mathit{sv}_1 } \}
  \end{align*}
  The overall effect of the constraints in $\WFConst_{\mathit{cir}}$
  is that a valid plan identifies a circuit in the graph.

\item $\WFConst_{\mathit{ham}}$ contains type-1
  entailment constraints that are specified using the binary relation
  $\mathit{distinct} \subseteq \WFUsers \times \WFUsers$:
  \[
     \mathit{distinct} = \{ ((u, i), (v, j)) \in \mathit{UV}
     \times \mathit{UV} \mid u \neq v \}
   \]
  Intuitively, two users are related by $\mathit{distinct}$ whenever
  they represent two distinct vertices.
  $\WFConst_{\mathit{ham}}$ can now be defined as follows:
  \[
      \WFConst_\mathit{ham} = \{ \ENT{ \mathit{distinct} }{
        \mathit{sv}_i }{ \mathit{sv}_j } \mid 1 \leq i < j
      \leq N \}
    \]
  Effectively, $\WFConst_{\mathit{ham}}$ ensures that any circuit
  identified by a valid plan passes through pairwise distinct
  vertices: i.e., the circuit is a Hamiltonian cycle.
 \end{enumerate}
\end{enumerate}
It is obvious that the reduction can be computed in time polynomial to
the size of the \DHC instance.
Observe also the following:
\begin{enumerate}
\item There is a one-to-one correspondence between a valid plan of
  \aWAP and a Hamiltonian cycle in $G$.
\item For every choice of $D \subseteq B$ with $|D| \leq |B|/2$, there
  is a corresponding set of no more than $t$ users from $\bigcup_{e
    \in B}\mathit{UE}_e$ that the adversary can remove.
\item When the adversary removes $t$ users from \WFUsers, no more than
  $t$ of them belong $\bigcup_{e \in B}\mathit{UE}_e$ .  These
  latter users correspond to a choice of $D \subseteq B$ with
  $|D| \leq |B|/2$.
\end{enumerate}
Consequently, the input instance $(G, B)$ belongs to \DHC if and only
if the output instance $(\aWAP, t)$ belongs to \SRCP.
\end{proof}
The reduction above employs only type-1 entailment constraints,
meaning that type-1 entailment is all that is required to drive the
complexity of \SRCP to the second level of the polynomial hierarchy.

\section{One-shot Resiliency}
\label{sec-orcp}

An impression that one may get from reading \cite{Wang-Li:2007,
  Wang-Li:2010} is that, with static resiliency as an exception, other
notions of workflow resiliency (such as decremental and dynamic
resiliency) are largely \PSPACE-complete because of their game-based
definition.  The goal of this section is to dispel this false
impression. We do so by proposing a notion of workflow resiliency that
is more sophisticated than static resiliency, and yet remains in the
lower levels of the polynomial hierarchy.

\subsection{Problem Definition}

  One-shot resiliency is a generalization of static resiliency and a
  specialization of decremental resiliency.  Rather than removing
  users at the beginning of the game, as in static resiliency, the
  adversary of one-shot resiliency may wait till a more opportune
  time, and then remove users in the middle of the game.  Yet, unlike
  decremental resiliency, in which the adversary may ``strike''
  multiple times, the adversary in one-shot resiliency may only strike
  at most once.
\begin{definition}[One-shot Resiliency]
  A workflow authorization policy $\aWAP = \WFSchema{ \WFSteps }{
    \WFPrec }{ \WFUsers }{ \WFSAuth }{ \WFConst }$ is \Dfn{one-shot
    resilient} for integer budget $t \geq 0$ if and only if Player 1
  can win the \Dfn{one-shot resiliency game} no matter how
  Player 2 plays.

  The one-shot resiliency game is a two-player game that proceeds in
  rounds. At any time, the configuration of the game is a pair
  $(\aRemSet, \aPPlan)$, where $\aRemSet \subseteq \WFUsers$, and
  $\aPPlan$ is a partial plan. Initially, $\aRemSet$ and \aPPlan are
  both $\emptyset$.

  Each round begins by Player 2 opting to either \Dfn{pass} or
  \Dfn{strike}, with the restriction that Player 2 must pass in all
  future rounds after it has struck in a round.  If Player 2 chooses
  to strike, then it further selects no more than $t$ users from
  \WFUsers to be placed in $\aRemSet$.  No action is required of
  Player 2 if it passes. Next, Player 1 extends \aPPlan by assigning a
  member of $\WFUsers \setminus \aRemSet$ to a not-yet-assigned step,
  so that \aPPlan remains causally closed.

  Player 2 wins right away if
  $\WFUsers \setminus \aRemSet = \emptyset$, or if \aPPlan becomes
  invalid.  Player 1 wins if \aPPlan is eventually extended to a valid
  plan.
\end{definition}
One-shot resiliency models situations in which the workflow runs for a
moderate length of time (as in decremental resiliency), but
a catastrophe is a truly rare event. The latter either does not occur,
or else it occurs only once.  The effect of the catastrophe is
irreversible during the execution of the workflow, as in the cases of
static and decremental resiliency (i.e., removed users do not return
to the game).

We write \ORCP to denote the language containing pairs $(\aWAP, t)$ so
that \aWAP is one-shot resilient for budget $t$.
\begin{theorem}
  $\SRCP \subset \ORCP \subset \CRCP$.
\end{theorem}
The proof of this theorem is left as an exercise for the reader.

Even though it is defined in terms of a strategic game, one-shot
resiliency remains in the third level of the polynomial hierarchy.
\begin{theorem} \label{thm-orcp-sigma-three-complete}
  \ORCP is \SigmaThree-complete.
\end{theorem}
A proof of this result will be given in \S \ref{sec-orcp-membership}
and \S \ref{sec-orcp-hardness}.  This result provides the intellectual
justification for deciding one-shot resiliency through a reduction to
first-order Answer-Set Programming with bounded predicate arities (\S
\ref{sec-asp-encoding})

\subsection{Membership in \SigmaThree}
\label{sec-orcp-membership}

\begin{algorithm}[t]
\KwIn{a workflow authorization plan $\aWAP = \WFSchema{ \WFSteps }{ \WFPrec }{ \WFUsers }{
  \WFSAuth }{ \WFConst }$.}
\KwIn{an integer budget $t \geq 0$.}
\KwOut{a boolean value indicating if \aWAP is one-shot resilient for
  budget $t$.}
\KwRem{This algorithm is equipped with an \SRCP oracle.}
Guess a functional sequence $\anotherSeq = (\aStep_1, \aUser_1)\cdot
  (\aStep_2, \aUser_1) \cdot \ldots \cdot (\aStep_{|\WFSteps|},
  \aUser_{|\WFSteps|})$\;
\lIf{$\aPPlan_\anotherSeq$ is not a valid plan}{\Return{false}}
$\aWAP_0 \leftarrow \aWAP$\;
\For{$i$ from $1$ to $|\WFSteps|$}{
  \lIf{$(\aWAP_{i-1}, t) \not\in \SRCP$\label{line-orcp-oracle}}{\Return{false}}
  $\aWAP_i \leftarrow \mathit{project}(\aWAP_{i-1}, \aStep_i, \aUser_i)$\label{line-orcp-project}\;
}
\Return{true}\;
\caption{A non-deterministic algorithm for deciding \ORCP.\label{algo-orcp}}
\end{algorithm}

We begin the proof of \SigmaThree-completeness by arguing that \ORCP
is in \SigmaThree. This argument turns out to be anything but trivial,
and it sheds light on the problem structure of \ORCP: there is a short
encoding of a winning strategy for Player 1. This insight will be used
in our ASP encoding of \ORCP in \S \ref{sec-asp-encoding}.

Suppose \aWAP is a workflow authorization policy \aWAP. We construct a
decision tree $\mathcal{T}_\aWAP$ that captures the ``moves'' of
Player 1.  A sequence $\aSeq \in (\WFSteps \times \WFUsers)^*$ is
\Dfn{functional} if and only if no step appears more than once in
\aSeq. Every functional sequence represents a function $\aPPlan_\aSeq$
that maps steps to users.  When $\aPPlan_\aSeq$ is a valid partial
plan, we call \aSeq a \Dfn{play} of Player 1.  Each play represents a
legitimate sequence of ``moves'' that Player 1 can make (without
losing).  The decision tree $\mathcal{T}_\aWAP$ is constructed as
follows: (a) tree nodes are plays; (b) a play \aSeq is the parent of
another play \anotherSeq whenever
$\anotherSeq = \aSeq \cdot (\aStep, \aUser)$ (i.e., \anotherSeq is
obtained from \aSeq by assigning a user to an additional step).  In
$\mathcal{T}_\aWAP$, the empty sequence $\epsilon$ is the root of the
tree, and a play \anotherSeq is a descendent of another play \aSeq
whenever \aSeq is a prefix of \anotherSeq.  A play \aSeq is called a
\Dfn{terminus} whenever $\aPPlan_\aSeq$ is a valid plan (i.e., every
step is assigned).  \aWAP is satisfiable if and only if
$\mathcal{T}_\aWAP$ has at least one terminus.

Suppose $t$ is the budget for Player 2. A \Dfn{strike} of Player 2 is
a pair of the form $(\aSeq, \aRemSet)$, where \aSeq is a play,
$\Dom{ \aPPlan_\aSeq } \neq \WFSteps$, $\aRemSet \subseteq \WFUsers$,
and $|\aRemSet| \leq t$. The play \aSeq is the \Dfn{trigger} of the
strike.  Intuitively, a strike is a rule that tells Player 2 to remove
the users in \aRemSet immediately after Player 1 has made the moves in
\aSeq.  A strike $(\aSeq, \aRemSet)$ \Dfn{invalidates} a terminus
\anotherSeq when (a) \anotherSeq is a descendent of \aSeq in
$\mathcal{T}_\aWAP$, and (b) no user from \aRemSet is assigned by
\anotherSeq after the moves in \aSeq.  A strike is \Dfn{successful} if
and only if it invalidates every terminus that is a descendent of its
trigger.  A set $\mathcal{S}$ of successful strikes is a
\Dfn{strategy} for Player 2 if the strikes are pairwise
\Dfn{independent}: two strikes are independent whenever they have
distinct triggers that are not descendents of each other.  The
requirement of independence ensures that Player 2 strikes at most once
during a game play.  A strategy $\mathcal{S}$ of Player 2 is a
\Dfn{winning strategy} if and only if every terminus of
$\mathcal{T}_\aWAP$ is invalidated by a strike in $\mathcal{S}$.  A
strategy $\mathcal{S}'$ \Dfn{subsumes} another strategy $\mathcal{S}$
if and only if (a) every terminus that $\mathcal{S}$ invalidates is
also invalidated by $\mathcal{S}'$, and (a) $\mathcal{S}'$ invalidates
at least one terminus that $\mathcal{S}$ does not invalidate.  A
strategy $\mathcal{S}$ is \Dfn{maximal} if and only if there it is not
subsumed by any other strategy.

Suppose \aWAP is one-shot resilient for budget $t$. Consider a maximal
strategy $\mathcal{S}$ for Player 2. Since $\mathcal{S}$ cannot be a
winning strategy, there is at least one terminus \anotherSeq that is
not invalidated by any strike in $\mathcal{S}$. Here is the crux of
the present argument: No strike can invalidate \anotherSeq, or else we
can construct another strategy $\mathcal{S'}$ that subsumes
$\mathcal{S}$, thereby contradicting the maximality of $\mathcal{S}$.
The play \anotherSeq can be seen as a succinct representation of a
winning strategy for Player 1.
\begin{lemma} \label{lemma-ORCP-ver-cond} $(\aWAP, t) \in \ORCP$ if
  and only if there is a sequence \anotherSeq of assignments such that
  (a) $\aPPlan_\anotherSeq$ is a valid plan, and (b) for every play
  \aSeq that is a prefix of \anotherSeq (i.e., \aSeq is an ancestor of
  \anotherSeq in $\mathcal{T}_\aWAP$), no strike with trigger \aSeq
  can be successful.
\end{lemma}
The idea of Lemma \ref{lemma-ORCP-ver-cond} can be translated into a
non-deterministic algorithm for deciding \ORCP, as depicted in
Algorithm \ref{algo-orcp}.  The algorithm begins by guessing a play
\anotherSeq that corresponds to a valid plan $\aPPlan_\anotherSeq$,
and then check that no prefix \aSeq of \anotherSeq can trigger a
successful strike.  An insight is that this latter check can be
achieved by invoking an \SRCP oracle against the ``remaining
workflow'' after the assignments in \aSeq are committed.  This notion
of the ``remaining workflow'' is formalized in the following
definition, which defines the notation used in line
\ref{line-orcp-project}.
\begin{definition}
  Suppose $\aWAP = \WFSchema{ \WFSteps }{ \WFPrec }{ \WFUsers }{
    \WFSAuth }{ \WFConst }$ is a workflow authorization policy, and
  $\{ (\aStep, \aUser) \}$ is a valid partial plan for \aWAP. Then
  $\mathit{project}( \aWAP, \aStep, \aUser )$ is the
  workflow authorization plan \WFSchema{ \WFSteps' }{ \WFPrec' }{
    \WFUsers' }{ \WFSAuth' }{ \linebreak \WFConst' } such that:
\begin{align*}
  \WFSteps' &= \WFSteps \setminus \{ \aStep \} \\
  \WFPrec' &= (\WFPrec) \cap (\WFSteps' \times \WFSteps') \\
  \WFUsers' &= \WFUsers \\
  \WFSAuth' &= \WFSAuth \cap (\WFSteps' \times \WFUsers')
\end{align*}
and $\WFConst'$ is defined as follows:
\begin{multline*}
  \WFConst' = \{ (\aStepSet, \aPPlanSet ) \mid
                  (\aStepSet, \aPPlanSet) \in \WFConst,
                              \aStep \not\in \aStepSet \} 
    \cup \mbox{}\\
\{ (\aStepSet \setminus \{ \aStep \}, \aPPlanSet\langle\aStep,
        \aUser\rangle ) 
     \mid (\aStepSet, \aPPlanSet) \in \WFConst,
          \aStep \in \aStepSet, |\aStepSet| > 1 \}
\end{multline*}
where the notation $\aPPlanSet \langle \aStep, \aUser\rangle$ is
defined below:
\[
\aPPlanSet \langle \aStep, \aUser\rangle =
    \{ \aPPlan \setminus \{ (\aStep, \aUser) \} \mid
       \aPPlan \in \aPPlanSet, (\aStep, \aUser) \in \aPPlan \}
\]
Intuitively, $\aPPlanSet \langle \aStep, \aUser \rangle$ selects those
partial plans in \aPPlanSet that are consistent with the assignment of
\aUser to \aStep, and then eliminates the pair $(\aStep, \aUser)$ from
those selected partial plans.
\end{definition}

Since \SRCP is \PiTwo-complete, Algorithm \ref{algo-orcp} depends on
an $\NP^\NP$ oracle. In addition, Algorithm \ref{algo-orcp} runs in
non-deterministic polynomial time. Therefore, \ORCP belongs to
\SigmaThree.

\subsection{\SigmaThree-hardness}
\label{sec-orcp-hardness}

To demonstrate that \ORCP is hard for \SigmaThree, we present a
polynomial-time Karp reduction from \SuccRad{k} to \ORCP. The problem
\SuccRad{k} is known to be \SigmaThree-complete for every $k \geq 2$
\cite{SkR, PHCatalog}.

The \Dfn{radius} of a directed graph $G = (V, E)$ is the smallest $k$
such that there exists a vertex $u \in V$ such that every vertex
$v \in V$ is reachable from $u$ by some directed path of length no
greater than $k$. (We consider $u$ reachable from itself by a directed
path of length zero.)  Deciding if $G$ has a radius no greater than
$k$ is not hard if $G$ is represented as, say, an access control
matrix or adjacency lists. What causes the problem to become
intractable is when $G$ is specified using a \Dfn{succinct
  representation}. In such a representation, the adjacency matrix of
$G$ is specified through a boolean circuit.  Suppose $V$ has a size of
$2^n$ for some $n \geq 1$, then vertices can be identified by bit
vectors of length $n$. The adjacency matrix of $G$ can now be
represented by a boolean circuit $\mathit{BC}_G$ that takes two
size-$n$ bit vectors $\vec{x}$ and $\vec{y}$ as input, and returns a
one-bit value to indicate if there is a directed edge from vertex
$\vec{x}$ to vertex $\vec{y}$.  Formulated in this way, deciding if
the radius of $G$ is bounded by $k$ is \SigmaThree-complete.
\begin{description}
\item[Problem:] \SuccRad{k} \cite{SkR}
\item[Instance:] A boolean circuit $\mathit{BC}_G$ that succinctly
  represents a directed graph $G$
\item[Question:] Is the radius of $G$ no greater than $k$?
\end{description}

Given $\mathit{BC}_G$, one can construct a boolean circuit
$\mathit{BC}_{\mathit{reach}}(\vec{x}, \vec{y} ; \vec{z}^1, \linebreak
\vec{z}^2, \ldots, \vec{z}^{k-1})$, which takes $k+1$ size-$n$ bit
vectors as input, and returns $1$ if and only if vertex $\vec{y}$ is
reachable from vertex $\vec{x}$ by a directed path of length no more
than $k$, and that directed path visits intermediate vertices $\vec{z}^1$,
$\vec{z}^2$, etc in that order.
\begin{multline*}
\mathit{BC}_{\mathit{reach}}(\vec{x}, \vec{y} ; \vec{z}^1, \vec{z}^2,
\ldots, \vec{z}^{k-1}) = 
   \mathit{BC}_=(\vec{x}, \vec{y}) \lor \mathit{BC}_G(\vec{x}, \vec{y}) \lor \mbox{}\\
   \mathit{BC}_{\mathit{walk}}(\vec{x}, \vec{y}; \vec{z}^1) \lor
   \ldots \lor \mathit{BC}_{\mathit{walk}}(\vec{x}, \vec{y}; \vec{z}^1, \ldots, \vec{z}^{k-1})
\end{multline*}
The circuit $\mathit{BC}_{=}(\vec{x}, \vec{y})$ tests if
$\vec{x}$ and $\vec{y}$ are identical bit vectors.  The circuit
$\mathit{BC}_{\mathit{walk}}$ is defined as follows:
\begin{multline*}
   \mathit{BC}_{\mathit{walk}}(\vec{x}, \vec{y}; \vec{z}^1, \vec{z}^2,
   \ldots, \vec{z}^i) = \mbox{}\\
  \mathit{BC}_G(\vec{x}, \vec{z}^1) \land 
  \mathit{BC}_G(\vec{z}^1, \vec{z}^2) \land 
  \ldots \land \mathit{BC}_G(\vec{z}^i, \vec{y})
\end{multline*}
Note that $\mathit{BC}_{\mathit{reach}}$ has a size that is $O(k^2)$
times the size of $\mathit{BC}_G$.

One can now check if $G$ has a radius no greater than $k$ by checking
the following:
\begin{equation} \label{eqn-orcp-quantification}
    \exists \vec{x} \,.\, \forall \vec{y} \,.\, \exists \vec{z}^{1},
    \vec{z}^{2}, \ldots, \vec{z}^{k-1} \,.\, \mathit{BC}_\mathit{reach} (\vec{x}, \vec{y}; \vec{z}^{1},
     \vec{z}^{2}, \ldots \vec{z}^{k-1}) = 1
\end{equation}
Our goal is now to encode \eqref{eqn-orcp-quantification} using an
\ORCP instance.

Our reduction takes as input an instance of \SuccRad{k} consisting of 
a boolean circuit $\mathit{BC}_G$, where $G$ has $2^n$ vertices, and constructs an instance of 
\ORCP consisting of a budget $t = n$ and a workflow authorization
policy $\aWAP = \WFSchema{ \WFSteps }{ \WFPrec }{ \WFUsers }{
    \WFSAuth }{ \WFConst }$:
\begin{enumerate}
\item The steps in \WFSteps are ``placeholders'' for boolean values
  representing the input and output bits of
  $\mathit{BC}_{\mathit{reach}}$, as well as the intermediate values
  computed by the gates in $\mathit{BC}_{\mathit{reach}}$.
\begin{itemize}
\item $\WFSteps = \WFSteps_x \cup \WFSteps_y \cup \WFSteps_z \cup
  \WFSteps_{\mathit{gate}} \cup \WFSteps_{\mathit{out}}$.
\item $\WFSteps_x = \{ x_1, \ldots, x_n \}$ and $\WFSteps_y  = \{ y_1,
  \ldots, y_n \}$ contain one step for each bit of $\vec{x}$ and $\vec{y}$. 
\item
$\WFSteps_z = \{ z_1^1, \ldots, z_n^1, z_1^2, \ldots, z_n^2,
  \ldots, z_1^{k-1}, \ldots, z_n^{k-1} \}$  contains one step for each bit
  in $\vec{z}^1$, $\vec{z}^2, \ldots, \vec{z}^{k-1}$, for a total of
  $n \times (k-1)$ steps.
\item
$\WFSteps_{\mathit{gate}}$ contains one step each gate in
  $\mathit{BC}_{\mathit{reach}}$. Intuitively, these steps represent
  the output bits of the gates.
\item $\WFSteps_{\mathit{out}} = \{ \mathit{out} \}$ contains exactly
  one step representing the output bit of $\mathit{BC}_{\mathit{reach}}$.
\end{itemize}
\item \WFPrec orders the steps in $\WFSteps_x$ first, then
  $\WFSteps_y$, followed by $\WFSteps_z$, and then
  $\WFSteps_{\mathit{gate}}$, and lastly $\WFSteps_{\mathit{out}}$.
\item Users represent the two boolean values (true and false).  A plan
  can thus be interpreted as an assignment of boolean values to the
  the input, output, and gates of the boolean circuit
  $\mathit{BC}_{\mathit{reach}}$.  Although there are only two boolean
  values, there are multiple copies for each.
\begin{align*}
\WFUsers &= \WFUsers_{\mathit{bool}} \cup \WFUsers_{\star} \\
\WFUsers_{\mathit{bool}} &= \WFUsers_\bot \cup \WFUsers_\top\\
\WFUsers_\bot &= \{ (\bot, 1), (\bot, 2), \ldots, (\bot, n+1) \}\\
\WFUsers_\top &= \{ (\top, 1), (\top, 2), \ldots, (\top, n+1) \}\\
\WFUsers_\star &= \{ (f, 1), (t, 2), (f, 3), (t, 4),
  \ldots, (f, 2n-1), (t, 2n) \}
\end{align*}
There are two variants of boolean values, the $\top/\bot$-variant
($\WFUsers_{\mathit{bool}}$), and the $t/f$-variant
($\WFUsers_\star$).  ``True'' is represented by $\top$- and $t$-users,
and ``false'' is represented by $\bot$- and $f$-users.  Each boolean
value of the $\top/\bot$-variant has $n+1$ copies (i.e., more than the
budget $t=n$).  There are, however, only $n$ copies of each boolean
value of the $t/f$-variant
\item The step authorization policy \WFSAuth describes what type of
  boolean values can be assigned to each step.
\begin{gather*}
  \WFSAuth = ((\WFSteps_x \cup \WFSteps_z \cup
  \WFSteps_{\mathit{gate}}) \times \WFUsers_{\mathit{bool}}) \cup\mbox{}\\
  (\WFSteps_{\mathit{out}}
  \times \WFUsers_{\top})  \cup
  (\WFSteps_y \times \WFUsers_\star) 
\end{gather*}
Essentially, boolean values of the $\top$/$\bot$-variant can be
assigned to steps representing $\vec{x}$, $\vec{z}^i$, and the circuit
gates.  The output of the entire boolean circuit is forced to be true,
as only $\top$-values can be assigned.  Only boolean values of the
$t$/$f$-variant can be assigned to steps representing $\vec{y}$.

\item $\WFConst = \WFConst_{\mathit{order}} \cup
  \WFConst_{\mathit{gate}} \cup \WFConst_{\mathit{out}}$ contains
  three types of constraint.:
  \begin{enumerate}

\item Constraints in $\WFConst_{\mathit{gate}}$ encode the computation
  performed by the gates in $\mathit{BC}_{\mathit{reach}}$.  For
  example, suppose step $\aStep \in \WFSteps_{\mathit{gate}}$
  corresponds to an AND gate, which in turn takes its two input bits
  from the output of the gates represented by steps
  $\aStep_1, \aStep_2 \in \WFSteps_{\mathit{gate}}$. Then we formulate
  a constraint \WFConstraint{ \{ \aStep, \aStep_1, \aStep_2 \} }{
    \aPPlanSet_{\mathit{and}} }, so that $\aPPlanSet_{\mathit{and}}$
  contains all partial plans \aPPlan such that
  $\aPPlan ( \aStep )$ is a user representing ``true'' if and only if
  both $\aPPlan ( \aStep_1 )$ and $\aPPlan ( \aStep_2 )$ are users
  represent ``true.''  Similar constraints can be formulated for OR
  gates and NOT gates.

\item $\WFConst_{\mathit{out}}$ contains exactly one type-1 entailment
  constraint \ENT{ \mathit{equal} }{ \mathit{out} }{ \aStep }, where
  $\aStep \in \WFSteps_{\mathit{gate}}$ represents the gate that
  computes the overall output of $\mathit{BC}_{\mathit{reach}}$, and
  $\mathit{equal} \subseteq \WFUsers \times \WFUsers$ is a binary
  relation such that $(\aUser_1, \aUser_2) \in \mathit{equal}$
  whenever $\aUser_1$ and $\aUser_2$ both represent the same boolean
  value. Since $\mathit{out} \in \WFSteps_{\mathit{out}}$ can only
  be assigned $\top$-users, this constraint forces
  $\mathit{BC}_{\mathit{reach}}$ to output ``true.''
  
\item $\WFConst_{\mathit{order}}$ contains a type-1 entailment
  constraints \ENT{ \mathit{order} }{ \linebreak y_i }{ y_{i+1} } for
  every pair of steps $y_i, y_{i+1} \in \WFSteps_y$. The binary
  relation
  $\mathit{order} \subseteq \WFUsers_\star \times \WFUsers_\star$ is
  defined in such a way that
  $((b_1, j_1), (b_2, j_2)) \in \mathit{order}$ if and only if
  $j_1 < j_2$.  These constraints forces a linearization of the
  $t$/$f$ truth values when they are assigned to $\vec{y}$.
\end{enumerate}
\end{enumerate}
The encoding of the boolean circuit $\mathit{BC}_{\mathit{reach}}$ is
straightforward to understand. We explain here how the quantification
structure ($\exists$-$\forall$-$\exists$) of formula
\eqref{eqn-orcp-quantification} is captured by the reduction. There
are $n+1$ copies of each boolean value of the $\top$/$\bot$-variant,
but the budget of Player 2 is only $t = n$. No matter which $t$ users
are removed by Player 2, Player 1 can freely assign any boolean values
of the $\top$/$\bot$-variant to the steps representing the bit vectors
$\vec{x}$ and $\vec{z}^i$.  That is not the case for $\vec{y}$.  If
Player 2 strikes before the steps in $\WFSteps_{\mathit{y}}$ are
assigned, and it also removes $t=n$ users of the $t$/$f$-variant, then
there are only $n$ such $t$/$f$ values left to be assigned to the $n$
steps in $\WFSteps_{\mathit{y}}$.  In addition, the constraints in
$\WFConst_{\mathit{order}}$ requires that the remaining $t$/$f$-users
are assigned to the $\WFSteps_{\mathit{y}}$ in ``sorted order'' of
their indices. This means that the remaining $n$ boolean values of the
$t$/$f$-variant are now linearized into a bit vector when they are
assigned to $\vec{y}$. In this way, Player 2 can dictate the value of
$\vec{y}$.  To maximize its control, Player 2 will (a) strike before
any of the $\WFSteps_y$-steps are assigned, (b) strike after all the
$\WFSteps_x$-steps are assigned so as to maximize its knowledge of
$\vec{x}$, and (c) remove only users from $\WFUsers_\star$ (since
removing users of the $\top$/$\bot$-variant has no effect on the
decisions of Player 1). The overall effect is that Player 1 will first
pick $\vec{x}$, then Player 2 picks $\vec{y}$, and after that Player 1
picks the $\vec{z}^i$'s.  This captures exactly the quantification
structure of the formula in \eqref{eqn-orcp-quantification}.

Now that we know \SRCP and \ORCP are respectively complete problems in
the second and third level of the polynomial hierarchy, we propose in
the following a solution approach for these two problems.

\section{Background: Answer-Set Programming}
\label{sec-asp}

Answer-Set Programming (ASP) is a declarative programming paradigm
\cite{ASP-Book}.  It is essentially Datalog with disjunction and
default negation, defined over a stable model semantics.  Over the
last decades, ASP implementations have become increasingly competitive
in efficiency.  A notable example of a mature ASP implementation is
the Potassco project \cite{Potassco}, which is the ASP solver used in
this work. Due to such progress in ASP implementation technologies,
reasoning tasks at the lower levels of the polynomial hierarchy have
now been regularly encoded in ASP (e.g., \cite{Brewka-etal:2017,
  Sakama-Rienstra:2017}).  To these computational problems, ASP plays
a role analogous to what \SAT is for the \NP-complete
problems.  

We offer here a brief introduction to the stable model semantics of
propositional ASP. This prepares the reader to understand how the
\Dfn{model saturation technique} \cite{Eiter-etal:1995} works in the
encodings of \S \ref{sec-asp-encoding}.

We begin with the abstract syntax of \emph{propositional} ASP.
An ASP program $P$ is a finite set of \Dfn{rules}.
A rule $r$ has the following form:
\[
  a_1 \lor \ldots \lor a_n \aspIf b_1, \ldots, b_k, \aspNot b_{k+1},
  \ldots, \aspNot b_m
\]
Here, the $a_i$'s and $b_i$'s are atoms (i.e., propositional symbols),
and at most one of $m$ or $n$ can be zero.  The \Dfn{head} of $r$,
written $H(r)$, is the set $\{ a_1, \ldots, a_n \}$, and the
\Dfn{body} of $r$ is
$B(r) = \{ b_1, \ldots, b_k, \linebreak \aspNot b_{k+1}, \ldots,
\aspNot b_m \}$.  We write $B^+(r)$ for $\{ b_1, \ldots, b_k \}$, and
$B^-(r)$ for $\{ b_{k+1}, \ldots, b_m \}$.  A rule with $m = 0$ is a
\Dfn{fact} (in which case we omit the ``\aspIf''). A rule with $n = 0$
is an \Dfn{integrity constraint}.  Intuitively, a rule asserts that
one of the head atom holds if the body, read as a conjunction of
literals, holds. A fact asserts a (disjunctive) condition
unconditionally. An integrity constraint asserts that the body does
not hold.

Unlike Prolog, whose semantics is defined procedurally, ASP is purely
declarative.  The semantics of ASP is defined in terms of stable
models \cite{Gelfond-Lifschitz:1988}. An interpretation $I$ of program
$P$ is a set of atoms.  $I$ is a \Dfn{model} of $P$ if and only if,
for every $r \in P$, $H(r) \cap I \neq \emptyset$ when
$B^+(r) \subseteq I$ and $B^-(r) \cap I = \emptyset$.  Not every model
is stable though.  To arrive at the definition of a stable model, we
need to define the \Dfn{reduct of $P$ relative to $I$}:
\[
  P^I = \{ H(r) \aspIf B^+(r) \,|\,
  r \in P,\, B^-(r) \cap I = \emptyset \}
\]
An interpretation $I$ is a \Dfn{stable model} of $P$ if and only if
$I$ is a $\subseteq$-minimal model of $P^I$. In other words, an
$I$ cannot be a stable model of $P$ if $P^I$ has a model $I'$
that is a proper subset of $I$.  This requirement of minimality is
crucial for understanding how the model saturation technique
\cite{Eiter-etal:1995} works in the ASP encodings of \S
\ref{sec-asp-encoding}.

The definition of stable models can be extended to first-order
programs through the use of Hebrand interpretations
\cite{Eiter-etal:2007}. The minimality requirement for the models of
the reduct carries to the first-order case.

Checking the existence of stable models in the presence of disjunction
and default negation is \SigmaTwo-complete for propositional ASP
\cite{Eiter-etal:1995}, and \SigmaThree-complete for first-order ASP
with bounded predicate arities \cite{Eiter-etal:2007}.  The complexity
results presented in \S \ref{sec-srcp} and \S \ref{sec-orcp} grant us
the rational justification for employing ASP technologies to solve
\SRCP and \ORCP.

\section{ASP Encoding of \SRCP and \ORCP}
\label{sec-asp-encoding}

ASP is a natural constraint solving technology for tackling problems
in the lower levels of the polynomial hierarchy.  Solving \SRCP and
\ORCP using ASP does not require us to assume that $t$ is a small
parameter (an assumption made in \cite{Crampton-etal:2017:JCS}).  This
section demonstrates the feasibility of this solution approach by
presenting ASP encodings of \SRCP and \ORCP instances.

The presentation below uses the concrete syntax of logic programs
supported by the Potassco collection of ASP solving tools
\cite{Potassco}. We do so to ensure realism: our encoding is literally
executable. This, however, does not affect the generality of our
encoding, as every Potassco-specific syntax employed by our encoding
can be reduced to pure ASP \cite[Chapter 2]{ASP-Book}.

In this section, only separation-of-duty constraints are encoded. The
encoding can be extended readily to accommodate binding-of-duty,
cardinality, and entailment constraints \cite{Crampton-etal:2012,
  Crampton-etal:2013}.  The focus here is not so much on the encoding
of various constraint types in ASP, but on using ASP to express the
quantification structure of \SRCP and \ORCP.

\subsection{Encoding Static Resiliency}
\label{sec-asp-srcp}

\begin{figure}
  \begin{tabular}{ccll} \hline
    \SRCP & \ORCP & \textit{Fact} & \\ \hline
 &&  \texttt{step(\aStep).} & \text{for each $\aStep\in\WFSteps$}\\
 \textit{ignored} &&  \texttt{before($\aStep_1$, $\aStep_2$).} &
         \text{whenever $\aStep_1 \WFPrecNE \aStep_2$}\\
 && \texttt{user(\aUser).} & \text{for each $\aUser\in\WFUsers$}\\
 && \texttt{auth(\aStep, \aUser).} &
            \text{for each $(\aStep, \aUser) \in \WFSAuth$} \\
 & \textit{ignored} & \texttt{sod($\aStep_1$, $\aStep_2$).} &
    for each $\SOD{\aStep_1}{\aStep_2} \in \WFConst$ \\ \hline
  \end{tabular}
\caption{Instance-specific facts used in the ASP encoding of 
  \SRCP and \ORCP. The predicate \texttt{before/2} is ignored in
  the \SRCP encoding, while \texttt{sod/2} is ignored in the \ORCP
  encoding.\label{fig-srcp-encoding-instance}}
\end{figure}

\newcounter{srcplinecount}
\newcommand{\srcpline}{\refstepcounter{srcplinecount}\textsc{\arabic{srcplinecount}}}
\newcommand{\srcplinelab}[1]{\srcpline\label{#1}}

\begin{figure}
\hrulefill
{\tt
\begin{tabbing}
xxxx\=xx\=\kill
\srcpline\>\% Generate Player 2's strike\\
\srcplinelab{line-srcp-gen-strike}\>\{ removed(U) : user(U) \} $t$.\\
\ \\
\srcpline\>\% Generate Player 1's assignment\\
\srcplinelab{line-srcp-avail}\>avail(S, U) :- auth(S, U), not removed(U).\\
\srcplinelab{line-srcp-assign}\>assign(S, U) : avail(S, U) :- step(S).\\
\ \\
\srcpline\>\% Test separation-of-duty constraints\\
\srcplinelab{line-srcp-violation}\>violation :- \\
\srcplinelab{line-srcp-violation-body}\>\>sod(S1, S2), assign(S1, U), assign(S2, U).\\
\ \\
\srcpline\>\% Model saturation\\
\srcplinelab{line-srcp-saturation}\>assign(S, U) :- violation, avail(S, U).\\
\ \\
\srcpline\>\% Reject unsaturated models\\
\srcplinelab{line-srcp-reject}\>:- not violation.
\end{tabbing}
}
\hrulefill
\caption{Rules common to all instances in the ASP encoding of
  \SRCP.\label{fig-srcp-encoding-common}}
\end{figure}

We present an ASP encoding for the \emph{complement} of \SRCP.  Given
a workflow authorization policy
$\aWAP = \WFSchema{ \WFSteps }{ \leq }{ \WFUsers }{ \WFSAuth }{
  \WFConst }$ and a budget $t \geq 0$, the pair $(\aWAP, t)$ belongs
to the complement of \SRCP if and only if there exists a subset
\aRemSet of no more than $t$ users such that every plan that does not
assign users from \aRemSet will fail to satisfy \aWAP.  In other
words, if our ASP encoding is unsatisfiable (has no stable model),
then \aWAP is statically resilient for budget $t$. Conversely, if our
ASP encoding is satisfiable (has at least one stable model), then
every stable model encodes a subset \aRemSet of users that can be
removed by the adversarial environment to render \aWAP unsatisfiable.

Our ASP encoding is \Dfn{uniform} \cite[Chapter 3]{ASP-Book} in the
sense that every instance of \SRCP can be reduced to a logic program
$P$ that can be ``factorized'' into two parts, a part $P_C$ containing
rules that are common to all instances of \SRCP, and an
instance-specific part $P_I$, such that $P = P_C \cup P_I$.

Suppose we have been given an \SRCP instance consisting of a workflow
authorization policy
$\aWAP = \WFSchema{ \WFSteps }{ \leq }{ \WFUsers }{ \WFSAuth }{
  \WFConst }$ and a budget $t \geq 0$.  The instance-specific part of
the ASP encoding consists of the facts in
Fig.~\ref{fig-srcp-encoding-instance}, which describe the components
\WFSteps, \WFPrecNE, \WFUsers, \WFSAuth, and \WFConst of \aWAP.  The
rules common to all instances of \SRCP are listed in
Fig.~\ref{fig-srcp-encoding-common}.

The rules on lines \ref{line-srcp-gen-strike}, \ref{line-srcp-avail},
and \ref{line-srcp-assign} in Fig.~\ref{fig-srcp-encoding-common} are
responsible for generating interpretations that serve as candidates
for stable models.

Line \ref{line-srcp-gen-strike} is a \Dfn{choice rule} that models the
choice of Player 2.  Specifically, it generates up to $t$ ground atoms
of the \texttt{remove/2} predicate in a model candidate. These facts
represent the set \aRemSet of users removed by Player 2.

Line \ref{line-srcp-avail} specifies when a user \aUser is available
for assignment to a step \aStep, given the choice of \aRemSet by
Player 2.  Line \ref{line-srcp-assign} is a shorthand that asserts,
for each step \aStep, the following disjunction: {\tt
\begin{tabbing}
xxxx\=\kill    
\>assign(\aStep, $\aUser_1$) $\lor$ $\ldots$ $\lor$
  assign(\aStep, $\aUser_m$).
\end{tabbing}
}
\noindent where $\aUser_1$, \ldots, $\aUser_m$ are the users identified
by predicate \texttt{avail/2} to be available for assignment to
\aStep.

The cumulating effect of the model generation rules (lines
\ref{line-srcp-gen-strike}, \ref{line-srcp-avail}, and
\ref{line-srcp-assign}) is that an interpretation
$I_{\aRemSet, \aPPlan}$ will be generated as a model candidate for
each user set \aRemSet of size $t$ or less, and for each plan \aPPlan
that both complies with the step authorization policy and assigns only
users from $\WFUsers \setminus \aRemSet$.  Each model candidate
$I_{\aRemSet, \aPPlan}$ contains, on top of the instance-specific
facts in Fig.~\ref{fig-srcp-encoding-instance}, the following ground
atoms:
\begin{itemize}
\item \texttt{removed(\aUser)} for each $\aUser \in \aRemSet$
\item \texttt{avail(\aStep, \aUser)} whenever \aUser is available for
  assignment to \aStep
\item \texttt{assign(\aStep, \aUser)} whenever $\aPPlan(\aStep) = \aUser$
\end{itemize}
There are now two cases:
\begin{description}
\item[Case 1:] \emph{\aPPlan is valid.}  Neither lines
  \ref{line-srcp-violation}--\ref{line-srcp-violation-body} nor line
  \ref{line-srcp-saturation} will be ``triggered.''  But line
  \ref{line-srcp-reject} will reject $I_{\aRemSet, \aPPlan}$.
\item[Case 2:] \emph{\aPPlan is not valid.}  Then lines
  \ref{line-srcp-violation}--\ref{line-srcp-violation-body} will
  detect this case, and introduce the proposition \texttt{violation}
  into the interpretation under consideration.  Moreover, the
  interpretation will then be \Dfn{saturated} by line
  \ref{line-srcp-saturation}: all possible ground atoms of predicate
  \texttt{assign/2} will be added to the interpretation.  In short,
  model candidate $I_{\aRemSet, \aPPlan}$ will be ``converted'' to a
  superset $I_{\aRemSet, *}$, which contains, on top of the ground
  atoms in $I_{\aRemSet, \aPPlan}$, the following ground atoms:
  \begin{itemize}
    \item \texttt{violation}
    \item \texttt{assign(\aStep, \aUser)} for every \aUser available
      for assignment to \aStep
  \end{itemize}
\end{description}

Suppose $(\aWAP, t) \in \SRCP$. For every choice of \aRemSet, there
exists at least one valid plan $\aPPlan_0$ that does not assign users
from \aRemSet. Note that the model candidate $I_{\aRemSet, \aPPlan_0}$
will not be saturated, and thus it will be rejected by line
\ref{line-srcp-reject}. Not only that, the saturated model
$I_{\aRemSet, *}$ is not a minimal model for $P^{I_{\aRemSet, *}}$,
because its proper subset $I_{\aRemSet, \aPPlan_0}$ is also a model
for $P^{I_{\aRemSet, *}}$.  Consequently, no stable model exists for
the program $P$.

Conversely, suppose $(\aWAP, t) \not\in \SRCP$. Then there is a
$\aRemSet_0$ such that every plan \aPPlan that does not assign users
from $\aRemSet_0$ is invalid.  This means that Case 1 above never
holds for this choice of $\aRemSet_0$.  This also means that every
model candidate $I_{\aRemSet_0, \aPPlan}$ is converted into a
saturated model $I_{\aRemSet_\star, *}$.  As none of the
interpretations $I_{\aRemSet_0, \aPPlan}$ is a model for
$P^{I_{\aRemSet_\star, *}}$, $I_{\aRemSet_\star, *}$ 
is the minimal model for $P^{I_{\aRemSet_\star, *}}$. The program $P$
has at least one stable model.

We have thus demonstrated that it is feasible to use ASP for encoding
\SRCP.  The key is to use an advanced ASP programming technique known
as model saturation \cite{Eiter-etal:1995} to encode the
quantification structure (i.e., $\exists$-$\forall$) of \SRCP's
complement.  We now examine an extension of this technique for
encoding \ORCP.

\subsection{Encoding One-shot Resiliency}
\label{sec-asp-orcp}

\newcounter{orcplinecount}
\newcommand{\orcpline}{\refstepcounter{orcplinecount}\textsc{\arabic{orcplinecount}}:}
\newcommand{\orcplinelab}[1]{\orcpline\label{#1}}

\begin{figure}
\hrulefill
{\tt
  \begin{tabbing}
xxxx\=xx\=\kill
\orcpline\>\% Generate a plan as part of Player 1's strategy\\
\orcplinelab{line-orcp-plan}\>1 \{ assign(S, U) : auth(S, U) \} 1 :- step(S).\\
\ \\
\orcpline\>\% Generate a total ordering of steps as part of \\
\orcpline\>\% Player 1's strategy\\
\orcplinelab{line-orcp-order-head}\>order(X, Y); order(Y, X) :-\\
\orcplinelab{line-orcp-order-body}\>\>step(X), step(Y), X != Y.\\
\orcplinelab{line-orcp-order-before}\>order(X, Y) :- before(X, Y).\\
\orcplinelab{line-orcp-order-trans}\>order(X, Z) :- order(X, Y), order(Y, Z).\\
\ \\
\orcplinelab{line-orcp-launch-comment}\>\% Generate strike point of Player 2\\
\orcplinelab{line-orcp-launch-choice}\>post(S); pre(S) :- step(S).\\
\orcplinelab{line-orcp-launch-consist}\>post(S2) :- post(S1), order(S1, S2).\\
\ \\
\orcplinelab{line-orcp-strike-comment}\>\% Generate strike set of Player 2\\
\orcplinelab{line-orcp-strike-remove}\>removed(U); preserved(U) :- user(U).\\
\ \\
\orcplinelab{line-orcp-avail-comment}\>\% Available assignments for Player 1\\
\orcplinelab{line-orcp-avail-pre}\>avail(S, U) :- pre(S), assign(S, U).\\
\orcplinelab{line-orcp-avail-post}\>avail(S, U) :- post(S), auth(S, U), preserved(U).\\
\ \\
\orcplinelab{line-orcp-inst-comment}\>\% Detect satisfiability\\
\orcplinelab{line-orcp-inst-head}\>sat :-\\
\orcplinelab{line-orcp-inst-gen}\>\>avail(1, U1), avail(2, U2), \textrm{\ldots}, avail(9, U9),\\
\orcplinelab{line-orcp-inst-test}\>\>U2 != U7, U3 != U4,
             \textrm{\ldots}, U8 != U9.\\
\ \\
\orcpline\>\% Player 2 loses if it removes more than $t$ users\\
\orcplinelab{line-orcp-budget}\>sat :- $t$+1 \{ removed(U) : user(U) \}.\\
\ \\
\orcpline\>\% Model saturation\\
\orcplinelab{line-orcp-saturate-first}\>pre(S) :- sat, step(S).\\
\orcpline\>post(S) :- sat, step(S).\\
\orcpline\>removed(U) :- sat, user(U).\\
\orcpline\>preserved(U) :- sat, user(U).\\
\orcplinelab{line-orcp-saturate-last}\>avail(S, U) :- sat, auth(S, U).\\
\ \\
\orcpline\>\% Reject unsaturated models\\
\orcplinelab{line-orcp-reject}\>:- not sat.
\end{tabbing}
}
\hrulefill
\caption{ASP encoding of \ORCP. Note that the rule on lines
  \ref{line-orcp-inst-head}--\ref{line-orcp-inst-test} are instance
  specific. This encoding also assumes the instance-specific facts in
  Fig.~\ref{fig-srcp-encoding-instance}.\label{fig-orcp-encoding-common}}
\end{figure}

The crux in designing an ASP encoding of \ORCP lies in capturing the
quantification structure of \SigmaThree (i.e.,
$\exists$-$\forall$-$\exists$).  To this end, we employ the advanced
adaption of model saturation as found in the proof of Lemma 6 and
Lemma 7 in \cite{Eiter-etal:2007}.

Unlike the encoding in \S \ref{sec-asp-srcp}, the ASP encoding of
\ORCP presented here is not uniform.  It cannot be factorized into an
instance-specific set of facts and a set of of rules that are common
to all instances.  More specifically, \ORCP is encoded using the
instance-specific facts in Fig.~\ref{fig-srcp-encoding-instance} and
the rules in Fig.~\ref{fig-orcp-encoding-common}. Note that lines
\ref{line-orcp-inst-head}--\ref{line-orcp-inst-test} of
Fig.~\ref{fig-orcp-encoding-common} are only examples. Each \ORCP
instance will have a different formulation of those lines, depending
on what constraints are in \WFConst.

Our ASP encoding of \ORCP basically follows the idea of Lemma
\ref{lemma-ORCP-ver-cond}, and thus
Fig.~\ref{fig-orcp-encoding-common} could be seen as an ASP variant of
Algorithm \ref{algo-orcp}:
\begin{enumerate}
  \item The logic program ``guesses'' a functional
    sequence \aSeq that encodes the strategy of Player 1.
  \item Then model saturation is employed for capturing the universal quantification
    of Player 2's strikes.
  \item Lastly, testing whether a prefix \anotherSeq of
\aSeq can be extended to a terminus is performed without model
generation rules.
\end{enumerate}

\paragraph{Generating Player 1's strategy.}
The first part of Player 1's strategy is a plan, which is represented
by the predicate \texttt{assign/2}. Using a choice rule, line
\ref{line-orcp-plan} generates, for each step \aStep, exactly one fact
of the form \texttt{assign(\aStep, \aUser)} if user \aUser is
authorized to execute step \aStep.

The second part of Player 1's strategy is a total ordering of steps.
This is represented by the predicate \Code{order/2}, which is
generated by lines
\ref{line-orcp-order-head}--\ref{line-orcp-order-trans}.  More
specifically, lines
\ref{line-orcp-order-head}--\ref{line-orcp-order-body} generate, for
every pair of distinct steps $X$ and $Y$, either \texttt{order($X$,
  $Y$)} or \texttt{order($Y$, $X$)}.  Line
\ref{line-orcp-order-before} ensures that the generated ordering
honors the ordering constraints of the workflow, and line
\ref{line-orcp-order-trans} ensures transitivity.

\paragraph{Generating Player 2's strike.}
The strike of Player 2 is generated by lines
\ref{line-orcp-launch-comment}--\ref{line-orcp-strike-remove}. There
are two parts to the strike. The first part is a subset \aRemSet of
users to be removed.  Line \ref{line-orcp-strike-remove} is a
shorthand that asserts, for each user \aUser, the disjunction below:
{\tt
\begin{tabbing}
xxxx\=\kill    
\>removed(\aUser) $\lor$ preserved(\aUser).
\end{tabbing}
}
\noindent The alert reader will notice that no constraint on the size
of \aRemSet is placed here. As we shall see, the size is controlled by
having Player 2 loses the game if it picks more than $t$ users in
\aRemSet (see line \ref{line-orcp-budget}).

The second part of Player 2's strike is the choice of a round to
launch the strike. This choice is generated by lines
\ref{line-orcp-launch-choice}--\ref{line-orcp-launch-consist}.  More
specifically, every step is labelled as either ``pre-strike'' or
``post-strike'' by line \ref{line-orcp-launch-choice}.  A pre-strike
step is one that is ordered before the launch point according to the
total ordering chosen by Player 1 above; otherwise, the step is
post-strike.  Line \ref{line-orcp-launch-consist} ensures that
post-strike steps are never ordered before pre-strike steps.

\paragraph{Testing if a prefix of Player 1's strategy can be extended
  to a terminus.} 
Lines \ref{line-orcp-avail-comment}--\ref{line-orcp-inst-test} check
whether, after the strike chosen by Player 2, the assignments made
prior to the strike by Player 1 can be extended to a valid plan.
This section can be further divided into two subsections: 
(a) lines \ref{line-orcp-avail-pre}--\ref{line-orcp-avail-post}, and
(b) lines \ref{line-orcp-inst-head}--\ref{line-orcp-inst-test}.

Lines \ref{line-orcp-avail-pre}--\ref{line-orcp-avail-post} determine
which user \aStep is available for assignment to which step \aStep
after the strike.  For pre-strike steps, the assignment is fixed
according to Player 1's strategy (line \ref{line-orcp-avail-pre}).
For post-strike steps, authorized users who have not been removed by
the strike are available (line \ref{line-orcp-avail-post}).

Then comes lines \ref{line-orcp-inst-head}--\ref{line-orcp-inst-test},
which mark the interpretation by the marker proposition \Code{sat}
whenever the workflow authorization policy is found to be satisfiable
(given Player 1's strategy and Player 2's strike).  This is the part
of the encoding that has been inspired by the proof of Lemma 7 in
\cite{Eiter-etal:2007}. The main feature of this section is that
satisfiability is checked without model generation rules (e.g., choice
rules and disjunctive rules).  Instead, a single rule is used. This
rule, however, is instance specific: a different rule of this form
will have to be formulated for each \ORCP instance.  The rule
considers all possible user assignments to the steps (line
\ref{line-orcp-inst-gen}), and then checks the SOD constraints inline
(line \ref{line-orcp-inst-test}).  By encoding the nested existential
quantification without model generation rules, we can reuse the model
saturation technique to encode the outer existential-universal
quantifications.

\paragraph{Other mechanics.}
Line \ref{line-orcp-budget} ensures that Player 2 plays according to
the budget. If Player 2 removes more than $t$ users, then Player 1
wins: i.e., the workflow authorization policy is considered satisfied.

If the workflow authorization policy is satisfiable (i.e., the
proposition \texttt{sat} is part of the model), then lines
\ref{line-orcp-saturate-first}--\ref{line-orcp-saturate-last} will
saturate the model by producing all possible ground atoms that can
ever be asserted as a result of Player 2's choice.  Finally, line
\ref{line-orcp-reject} rejects unsaturated models.  Using an argument
analogous to the one in \S \ref{sec-asp-srcp}, one can
demonstrate that a stable model exists if and only if Player 1 has a
winning strategy.

\section{Related Work}

The notion of resiliency was introduced into the study of access
control by Li \emph{et al.}, originally in the context of RBAC rather
than in workflow authorization models \cite{Li-etal:2006,
  Li-etal:2009}. Significant recent advances have been achieved in
employing parameterized complexity analysis to facilitate the design
of efficient algorithms for the Resiliency Checking Problem (RCP)
\cite{Crampton-etal:2016}.  In particular, the efficient FPT
algorithms designed for WSP is employed as subroutines for solving
RCP.  There is also interest in formulating ``resiliency''-variants of
combinatorial problems in general \cite{Crampton-etal:2017:CIAC}.  A
form of resiliency checking problem has also been defined for
Relationship-Based Access Control (ReBAC), in the context of policy
negotiation performed among co-owners of the a resource \cite{Pooya}.
ReBAC resiliency checking was shown to be complete for \PiTwo.  That
result, however, is significantly different from Theorem
\ref{thm-srcp-pi-two-hard} of the present paper. First, ReBAC
resiliency is about an adversary who can remove relationship edges in
an underlying social graph, whereas the adversary of workflow
resiliency removes users. Second, the hardness proof of \cite{Pooya}
involves a reduction from the Graph Consistency problem, and is
therefore fundamentally different from the one presented in this
paper.  In general, the nature of RCP is different from workflow
resiliency, as the latter contains an element of dynamism, allowing
the adversarial environment to remove users as the workflow is
executing.

The idea of workflow resiliency was first introduced by Wang and Li
\cite{Wang-Li:2007, Wang-Li:2010}. The present paper bridges the
complexity gap of static resiliency that existed since
the first conception of the idea a decade ago.

Two notable lines of recent research explore new directions in the
study of workflow resiliency.  Mace \emph{et al.}
\cite{Mace-etal:2014, Mace-etal:2015:QEST, Mace-etal:2015:SERENE}
pointed out that simply knowing whether a workflow is resilient or not
is not sufficient. The real interest of the workflow developer is to
receive feedback from the policy analysis in order to repair the
workflow. They proposed a quantitative model of resiliency in response
to this need. The idea is to build a probabilistic adversary model,
and then formulate a Markov Decision Process to capture how users are
removed over time. While the insight of their perspective is
acknowledged, it is the position of this paper that multiple notions
of resiliency are needed in order to provide insights into the
workflow engineering process.  The proposal of one-shot resiliency is
partly motivated by this consideration.

A second recent work on workflow resiliency is the FPT algorithm of
Crampton \emph{et al.} for deciding dynamic resiliency
\cite{Crampton-etal:2017:JCS}.  Unfortunately, the parameter they used
involves the budget $t$ of the adversary, which is not universally
considered small.  It is therefore necessary to consider solution
approaches that do not assume a small $t$. The proposal of ASP as a
solution approaches for static and one-shot resiliency in the present
work is a response to this challenge.

Kahn and Fong proposed \Dfn{workflow feasibility} as a dual notion of
workflow resiliency \cite{Khan-Fong:2012}. The idea is based whether
the current protection state can be repaired to make the workflow
satisfiable.  Feasibility checking is also defined for ReBAC
\cite{Pooya}, and shown to be complete for \SigmaTwo.

\section{Conclusion and Future Work}
\label{sec-conclusion}

We proved that static resiliency is complete for \PiTwo, thereby
solving a problem that has been open for more than a decade since Wang
and Li first proposed the notion of workflow resiliency
\cite{Wang-Li:2007, Wang-Li:2010}.

We have also demonstrated that useful notions of workflow resiliency
need not be \PSPACE-hard. The fact that we can define a notion of
workflow resiliency (one-shot resiliency) that remains in the lower
levels of the polynomial hierarchy (\SigmaThree) implies that the
lower complexity of static resiliency is not an exception.

The completeness of \SRCP and \ORCP in the second and third level of
the polynomial hierarchy also suggests that Answer-Set Programming is
a natural choice of constraint-solving technology for solving the two
problems (without having to assume that $t$ is a small parameter).  We
have demonstrated the feasibility of this approach by presenting ASP
encodings of \SRCP and \ORCP. These encodings involve the application
of the model saturation technique \cite{Eiter-etal:1995} and its
advanced adaptation \cite{Eiter-etal:2007}.

A number of research directions are suggested below:
\begin{enumerate}
\item It is unlikely that the ASP encodings of \SRCP and \ORCP
  presented in \S \ref{sec-asp-encoding} are the most efficient
  ones. Since the encodings are formulated mainly for demonstration of
  feasibility, they are optimized for brevity.  It is well-known that
  the efficiency of ASP programs can benefit from advanced
  optimization techniques \cite{ASP-Book}.  A natural follow-up work
  is to explore these optimization techniques, and empirically
  benchmark the performance of ASP-based solutions for \SRCP and
  \ORCP.  A promising direction is to combine the pattern-based
  technique of \cite{Cohen-etal:2014, Cohen-etal:2016} with ASP-solving.
  
  \item The proposal of \ORCP affirms the possibility for defining
    alternative notions of workflow resiliency beyond the three
    advanced by Wang and Li.  \ORCP not only avoids
    the \PSPACE-completeness of \CRCP and \DRCP, but also admits
    encoding in first-order ASP with bounded predicate arities.  An
    interesting research direction is to explore if there are other
    useful formulations of workflow resiliency that also reside in the
    lower levels of the polynomial hierarchy.
  \item The source of complexity for \PSPACE-complete notions of
    workflow resiliency (i.e., \CRCP and \DRCP) is that the adversary
    model is overly powerful. An open question is whether there are
    alternative adversary models for workflow resiliency that involve
    a less powerful adversary and still lead to useful notions of
    availability for workflow authorization models.  (In some sense,
    the quantitative model of Mace \emph{et al.} could be seen as an
    alternative adversary model \cite{Mace-etal:2014,
      Mace-etal:2015:QEST, Mace-etal:2015:SERENE}.) The existing
    adversary model is framed in terms of a budget $t$, and a flexible
    consumption schedule of this budget. Alternative adversary models
    can envision a different way in which the adversarial environment
    interacts with workflow execution.  Consider, for example, a
    \Dfn{small-accidents} variant of decremental resiliency, in which
    at most one user may be removed in each round. The assumption that
    $k + t$ is small now makes a lot of sense in this adversary model,
    and the FPT algorithm of \cite{Crampton-etal:2017:JCS} is
    genuinely reasonable in this model.  Creative deviation from the
    adversary model of Wang and Li is therefore a promising research
    direction.
  \item
    FPT-reductions to \SAT may be possible for
    some notions of workflow resiliency.  It is now a standard
    problem-solving technique to seek FPT reductions of hard problems
    to \SAT, so as to exploit the efficiency of \SAT-solving
    technologies \cite{deHaan-Szeider:KR2014, deHaan-Szeider:2017}.
    Bounded Model Checking is often cited as an example of this
    approach.  An even more general approach is to consider FPT
    algorithms that query a \SAT-solver multiple times (i.e., FPT
    Turing reductions) \cite{deHaan-Szeider:SAT2014}.  An open
    question is whether static and one-shot resiliency can be solved
    via FPT reductions to \SAT, and if not, whether there are other
    useful notions of workflow resiliency that can be solved in this
    way.
\end{enumerate}

\bibliographystyle{ACM-Reference-Format}


\end{document}